# Geometric spin dephasing of holes in two-dimensional nanostructures.

Yuri A. Serebrennikov

Qubit Technology Center

2152 Merokee Dr., Merrick, NY 11566

ys455@columbia.edu

We will show that the rate of geometric spin dephasing of a light hole (LH) localized in a two-dimensional (2D) nanostructure is equal to the rate of one-dimensional orientational relaxation of its crystal momentum. In contrast, geometric spin dephasing is singularly ineffective for heavy holes (HH) at zero magnetic field.

72.25.Rb  03.65.Vf  03.67.Lx

## I. Introduction

Spin dephasing destroys the entanglement in a spin subsystem, which is vital for quantum-logical operations. For that reason, considerable interest has been devoted to aspin relaxation of electrons and holes localized in a semiconductor nanostructures, where the reduced size and dimensionality lead to suppression of this process[1,2]. In semiconductors, the spin-orbit coupling (SOC) is rather strong and provides the main channel of coupling between the spin and a nonmagnetic stochastic "bath". Hence, design of the of solid-state quantum computers where the spin rather than the charge of electron is used for information processing and storage requires a thorough understanding of the SOC mediated relaxation.

Notably, most spin-orbit mechanisms of electron[3] and hole[4] spin-lattice relaxation in semiconductor nanostructures considered so far yield no decay of spin coherence in zero magnetic fields. In fact, it is commonly accepted that at zero magnetic field, $B = 0$, the time-reversal symmetry of a system prevents direct transitions between Kramers degenerate states even in the presence of SOC. It might therefore seem that at $B = 0$ adiabatic isolation of the Kramers doublet will make the SOC mediated spin relaxation ineffective. This assumption, however, is generally not correct.

The Bloch spinor depends on the orientation of the wave vector $\vec{k}$. The presence of intrinsic SOC caused by the atomic cores in the crystal couples *s*-like conduction band to the *p*-like valence band functions thereby mixing "spin-up" and "spin-down" electron Bloch functions[5] with the same lattice momentum. Due to this mixing an adiabatic and coherent change in the direction of a crystal momentum will lead to spin rotation and associated zero-field splitting (ZFS), which can be expressed as a manifestation of generally non-Abelian gauge field in the momentum space[6,7,8,9]. It has been shown[10] that at $B = 0$, an adiabatic (elastic) Coulomb scattering of $\vec{k}$ will lead to the random acquisition of geometric phases by the components of the Bloch spinor and, hence, to nonzero rate of spin decoherence. Due to a weak SOC in the lowest conduction band, this "geometric dephasing" yields relatively small spin relaxation rate described by the Elliott relationship $1/T_s \sim (\lambda/\Delta E)^2 \tau_k^{-1}$, where $\tau_k$ is the relaxation time of the orientation of the wave vector. The applicability of the Elliott relationship is limited by the constrain $\lambda/\Delta E \ll 1$, where $\Delta E$ is the interband energy separation and $\lambda$ characterize the amplitude of the interband matrix element of SOC. This condition is well satisfied for *s*-like electrons in the lowest conduction-band, but is not applicable for carriers with strong

SOC and/or near degenerate bands. Although, the geometric approach to spin relaxation is very general and is applicable for systems with arbitrary strong SOC, the consideration in Ref.[10] was focused on the electron spin relaxation.

Very recently, the decoherence of the hole angular momentum, $\vec{J}$, in bulk crystals was described within the framework of non-Markovian stochastic theory[11]. It has been shown that in the limit of strong SOC the adiabatic scattering of a hole wave vector does not lead to dephasing of the split-off holes, $1/T_J(\Gamma_7) \to \infty$. On the other hand, the same adiabatic relative to the $\Gamma_7$-$\Gamma_8$ splitting perturbation may scramble entanglement inside the $\Gamma_8$ quadruplet. Moreover, in the fast-motional limit one cannot distinguish between the HH and the LH bands. Note, however, that this prediction requires unrealistically short $\tau_k$ to wash out a LH-HH splitting $\Delta E_{HL}$ larger than 40 meV. In fact, the distinct optical orientation and relaxation of HH's and LH's that was clearly observed by Hilton and Tang[12] in undoped bulk GaAs at room temperature cannot be described by the theory presented in Ref.[11]. Situation becomes even more interesting for a hole localized in a 2D semiconductor nanostructure[13][14], where the size quantization leads to a large separation between the discrete energy levels $\Delta E_{HL} > 40$ meV, and the perturbation of the corresponding $LH_1$ and $HH_1$ bands by the $\vec{k} \to \vec{k}'$ scattering can be approximated as adiabatic, $\Delta E_{HL} \tau_k \gg 1$. Consequently, in line with the common expectation, suppression of inelastic scattering should inhibit the hole spin decoherence in 2D nanostructures. Yet, the question arises - whether the geometric dephasing that determines the fundamental *lower* limit[15][16] of spin decoherence in the conduction band will be ineffective for *holes* at $B = 0$. We address this question here.

We will show that the geometric Berry-phase shift acquired by the components of a LH spinor during an adiabatic scattering event $\vec{k} \to \vec{k}'$ is *equal* to the angular distance, $\vartheta_{\vec{k},\vec{k}'}$, traveled by $\vec{k}$. Consequently, stochastic reorientations of the hole crystal momentum induced by the anisotropic part of the hole-lattice or hole-impurity Coulomb interaction may lead to a very fast geometric dephasing. For a LH localized in a 2D nanostructure, the rate of this process is equal to the rate of one-dimensional orientational relaxation of the crystal momentum. Since the valence band has *p*-symmetry and does not couple through Fermi contact interaction to the nuclear spins (*s-p* hybridization of hole orbitals is clearly a second order effect), geometric dephasing is expected to be a dominant source of spin decoherence in LH's localized in 2D quantum dots at *B* = 0. We will also show that in a sharp contrast to light holes this mechanism is ineffective for heavy holes.

**II. Luttinger Hamiltonian in the rotating reference frame and geometric dephasing.**

Within the "spherical approximation"[17][18] the 4x4 matrix of the instantaneous Luttinger Hamiltonian can be expressed in the following form[11] (columns below correspond to m = 3/2, ½, -1/2, -3/2)

$$<m_1;k|H_{3/2}^{(M)}|m;k> = \frac{\gamma_1 k^2}{2m_0}\delta_{mm_1} + \frac{\gamma_2}{m_0}\begin{pmatrix} D_k & 0 & \sqrt{3}E_k & 0 \\ 0 & -D_k & 0 & \sqrt{3}E_k \\ \sqrt{3}E_k & 0 & -D_k & 0 \\ 0 & \sqrt{3}E_k & 0 & D_k \end{pmatrix} \quad (1)$$

Here the coefficients $\gamma_1$ and $\gamma_2$ are the dimensionless Luttinger parameters[17], $m_0$ is the bare electron mass, $D_k := -(2k_{z_M}^2 - k_{x_M}^2 - k_{y_M}^2)/2$, $E_k := -(k_{x_M}^2 - k_{y_M}^2)/2$. The superscript (*M*) denotes the *principal-axes* system of the "quadrupole" tensor $\vec{\vec{Q}}_{ij} = [L_i L_j + L_j L_i - (2L^2/3)\delta_{ij}]/2$, where $i, j = x_L, y_L, z_L$ represent the Cartesian basis in the space-fixed lab ($L$) frame, and $\vec{L}$ ($L = 1$) is the effective orbital angular momentum operator. This matrix can be can be expressed in terms of the irreducible tensor operators of the full rotation group:

$$H_{3/2}^{(M)} = (\gamma_1/2m_0)k^2 + (6)^{1/2}(\gamma_2/m_0)\sum_q (-1)^q K_{2q}^{(M)} T_{2-q}^{(M)}(J), \qquad (2)$$

where $T_{2q}(J) = (5/4)^{1/2}\sum_{mm_1} C_{Jm\,2q}^{Jm_1} |Jm_1><Jm|$, $J = L + S = 3/2$, $C_{1\mu 1\mu_1}^{2q}$ is the Clebsch-Gordon coefficient[19], and $K_{20}^{(M)} = (2/3)^{1/2} D_k$, $K_{2\pm1}^{(M)} = 0$, $K_{2\pm2}^{(M)} = E_k$. The last term on the RHS of Eq.(2), a scalar product of two spherical tensors of rank-2, reflects the coupling between $\vec{J}$ and the lattice momentum of a hole, and is clearly anisotropic. In the axially symmetric case ($E_k = 0$) the matrix of $H_{3/2}^{(M)}$ is diagonal in the $|LS, Jm>$ basis, $J_{z_M}$ is conserved, and the eigenfunctions of $H_{3/2}^{(M)}$ can be classified by the helicity $m = \hat{\vec{k}} \cdot \vec{J}^{(M)}$. Note that for holes moving along $z_M$ ($E_k = 0$, $D_k = -k^2$), Eq.(2) can be represented in the more familiar form $H_{3/2}^{(M)} = (k^2/2m_0)[\gamma_1 + 2\gamma_2(J^2/3 - J_{Z_M}^2)]$.

In the Elliott's mechanism of spin relaxation[6] the loss of coherence occurs only in the short time intervals during collisions. To describe the evolution of the system throughout the particular scattering event it is convenient to transform the basis into the moving frame of reference that coincides with the principal axis system of $\vec{\vec{Q}}_{ij}$ and follows the rotation of $\vec{k}$, $\Psi^{(M)}(t) = R^{(L)}(t)\Psi^{(L)}(t)$, where $\Psi$ is the instantaneous eigenvector of the *nontruncated* Luttinger Hamiltonian of the system. In our gauge convention the quantization axis of the system is chosen along the direction of the hole lattice momentum and corresponds to $z_M$ axis of the rotating *M*-frame. If the rotation of $\vec{k}$ is uniform, the operator $R^{(L)} = \exp[i\vec{\omega}^{(L)} \vec{J}^{(L)} t / \hbar]$ maps the *L*-frame into the actual orientation of the rotating *M*-frame at instance *t*. Here $\vec{\omega}^{(M)} = (d\vartheta_{\vec{k},\vec{k}'} / dt) \hat{\vec{n}}^{(M)}$ is the instantaneous angular velocity of the *M*-frame ($\omega_{x_M} dt = -\sin\theta_M \, d\varphi_M$, $\omega_{y_M} dt = d\theta_M$, $\omega_{z_M} dt = \cos\theta_M \, d\varphi_M$). Spherical angles, $\theta_M$ and $\varphi_M$, specify the orientation of the axis of rotation $\hat{\vec{n}}^{(M)}$ referred to the *M*-frame. In the presence of SOC the Bloch function is not factorizable into the orbital and spin parts, hence, the total angular momentum, $\vec{J} = \vec{L} + \vec{S}$, is a generator of the rotation $R^{(L)}$. In the rotating *M*-frame,

$$\widetilde{H}_{3/2}^{(M)}(t) = R^{(L)} H_{3/2}^{(L)}[\vec{k}(t)] R^{(L)-1} - i\hbar R^{(L)} \dot{R}^{(L)-1} = H_{3/2}^{(M)}[|\vec{k}(t)|] - \vec{\omega}^{(M)} \cdot \vec{J}^{(M)}, \quad (3)$$

The last term on the RHS of Eq.(3) represents the combined effect of the Coriolis and centrifugal terms, $-\vec{\omega}^{(M)} \vec{L}^{(M)}$, and spin-rotation interaction, $-\vec{\omega}^{(M)} \vec{S}^{(M)}$, as seen by an observer in the noninertial *M*-frame[20]. It is of interest to note that Eq.(3) is valid even if $\vec{k}$ rotation is not uniform (see Ref.[20]). Notably, as follows from Eqs.(1), (2), and (3) in the rotating *M*-frame the anisotropic part of $\widetilde{H}_{3/2}^{(M)}$ is formally equivalent to the expression for the effective *nuclear* quadrupole Hamiltonian of spin $I = 3/2$ in the rotating frame[21]. This is no coincidence since the symmetry group of both Hamiltonians is the same[22].

The next step entails adiabaticity of $\vec{k}$ rotation and leads to nontrivial gauge potentials and associated holonomies. Physically adiabaticity of the scattering event means that the inverse dwell time of the collision, $\sim 1/\tau_c$, is much smaller than the energy separation between bands. In other words, it requires vanishing interband transitions during the scattering event (elastic collisions), $\Delta E_{LH} \gg 1/\tau_c$, which is the usual approximation in the transport theory[23]. Therefore, if $|D_k| \gg |\vec{\omega}|$ $(\Delta E_{LH} \sim |D_k|, 1/\tau_c \sim |\vec{\omega}|)$, we may neglect the nondiagonal matrix elements of the rotating frame Hamiltonian, Eq.(3), and obtain adiabatically isolated Kramers doublets representing HH and LH, $\tilde{H}_{3/2}^{(M)} = \tilde{H}_{HH}^{(M)} + \tilde{H}_{LH}^{(M)}$. Projection of Eq.(3) onto the rotating basis spanned by the HH and LH spinor eigenfunctions of $H_{3/2}^{(M)}$ yields[24] (see also Ref.[23])

$$\tilde{H}_{HH}^{(M)} = (\gamma_1 - 2\gamma_2)(k^2/2m_0) + H_{HH,SR}^{(M)}, \quad H_{HH,SR}^{(M)} = -(3/2)\omega_{z_M}\sigma_{z_M}, \tag{4}$$

$$\tilde{H}_{LH}^{(M)} = (\gamma_1 + 2\gamma_2)(k^2/2m_0) + H_{LH,SR}^{(M)}, \quad H_{LH,SR}^{(M)} = -(1/2)\vec{\omega}^{(M)}\vec{\tilde{\gamma}}_{LH}^{(M)}\vec{\sigma}^{(M)}, \tag{5}$$

where $\vec{\sigma}$ is the vector of Pauli matrices and the "tensor" $\vec{\tilde{\gamma}}_{LH}^{(M)}$ in Eq.(5) is defined by the expression[25][26]

$$1/2\, \vec{\tilde{\gamma}}_{LH}^{(M)}\vec{\sigma}^{(M)} := P_{LH}^{(M)} \vec{J}^{(M)} P_{LH}^{(M)}, \tag{6}$$

$P_{LH}^{(M)} = |J1/2\rangle\langle J1/2| + |J-1/2\rangle\langle J-1/2|$, $\tilde{\gamma}_{LH;X_MX_M} = \tilde{\gamma}_{LH;Y_MY_M} = 2; \tilde{\gamma}_{LH;Z_MZ_M} = 1$.

Obviously, $\vec{\tilde{\gamma}}_{LH}^{(M)}$ is not a true tensor, since it does not transform covariantly under a gauge transformation into the rotating *M*-frame. It depends on a choice of gauge that specifies the *reference* orientation, i.e. the orientation in which the rotating *M*-frame coincides with some space-fixed frame. At the moment $t = 0$, this orientation may always be chosen (*locally* in the $\vec{k}$-space) such that $\vec{\tilde{\gamma}}_{LH}^{(M)}$ and $\vec{Q}^{(M)}$ are simultaneously diagonal, and that the direction of $\vec{k}$ coincides with the main axis $z_M$ which represents the quantization axis of the effective spin operator $\vec{\sigma}/2$.

The effective spin Hamiltonians on the RHS of Eqs.(4) and (5) have the form of a spin-rotation interaction and can be viewed as a generic Zeeman interaction of a spin-1/2 particle with an "effective" magnetic field that appears in the frame that follows the rotation of $\vec{k}$. The geometric nature of these results becomes clear if we take into account that a differential *action* of this field is proportional to the angle of rotation, $|\vec{\omega}(t)|\delta t$, i.e., to the *distance* in the angular space. As long as $\vec{k}$ rotation represents an adiabatic perturbation to the system, the evolution of the Bloch spinor is a unique function of the curve traversed by the lattice momentum in the angular space and is independent of the rate of traversal. Clearly, the non-Abelian gauge structure is present only in the LH band. Moreover, $[S_{Z_M}, H_{HH,SR}^{(M)}] = 0$, hence, the $z_M$ component of an effective spin operator $S_{Z_M} = \sigma_{Z_M}/2$ that represents the HH spinor is conserved and is not affected by an adiabatic rotation of the wave vector, $S_{Z_M} = S_{Z_L}$. Geometrically this result reflects the splitting of the corresponding Berry connection for states with the helicity difference $\Delta m > 1$, see Ref.[25] for details. This behavior is evidently the consequence of the second order approximation made in the $\vec{k} \cdot \vec{p}$ perturbation theory that leads to the Luttinger Hamiltonian[27].

If we assume that the plane of the $\vec{k}$ rotation remains constant, like in a strict 2D nanostructures, then the axis of rotation is normal to the lateral plane and can be assigned to say $x_M = x_L$ ($\theta_M = \pi/2$, $\varphi_M = 0$). In this case the corresponding Wilczek-Zee gauge potential[25][26][28], $A_{LH}^{(M)} = iH_{LH,SR}^{(M)}$, lost its non-Abelian character and Eq.(5) takes the form

$$\widetilde{H}_{LH}^{(M)} = (\gamma_1 + 2\gamma_2)(k^2/2m_0) - \omega_{x_L}\sigma_{x_L}. \tag{7}$$

To describe the evolution of the polarization vector, $\vec{u}_{LH}^{(L)}(t) := Tr[\rho_{LH}^{(L)}(t)\vec{S}^{(L)}]$, where $\rho_{LH}^{(L)}(t) := |\Psi_{LH}^{(L)}(t)><\Psi_{LH}^{(L)}(t)|$ is the density operator that represents the LH Kramers doublet in the local inertial reference frame we have to perform a reverse rotation of the basis compensating for the rotation of the *M*-frame, thereby closing the path in the angular space by the geodesic. This transformation is not associated with a physical change of a state and does not affect the kinetic energy of the carrier. In the 2D projective spinor-space, it is merely a $R_{KD}^{(L)-1} = \exp[-i\vec{\omega}^{(L)}\vec{\sigma}^{(L)}t/2]$. Taking into account Eq.(7), it is straightforward to obtain the following result (see Refs. [10, 24] for details):

$$u_{LH,z_L}(t) = Tr\{\sigma_{z_L}\exp[(-i\omega_{x_L}t/2)\sigma_{x_L}^\times]\sigma_{z_L}\} = \cos(\omega_{x_L}t) = \cos(\vartheta_{\vec{k},\vec{k}'}), \qquad (8)$$

where $\sigma_{x_L}^\times := [\sigma_{x_L},\ ]$. Thus, during a "slow" scattering event ($\Delta E_{HL} >> \tau_c^{-1}$), the effective spin of a LH will rigidly (adiabatically) follow the rotation of the crystal momentum. In the context of the problem under consideration Eq.(8) is only applicable during the collision ($t \leq \tau_c, \omega t \leq \pi$), in the local reference frame that reflects the geometry of the particular scattering event. Nevertheless, it clearly shows that the geometric Berry-phase shift acquired by the components of the LH-spinor during an elastic collision is equal to the angular distance traveled by $\vec{k}$, which reveals the geometric character of the phenomenon. Elastic scattering of a wave vector by fluctuations of the anisotropic part of a hole-lattice or hole-impurity Coulomb interaction will result in random phase shifts and lead to geometric dephasing of the effective spin, which represents the LH Kramers doublet[10,29,30,31].

As already mentioned, for a hole localized in a 2D nanostructure, the axis of rotation of the local reference frame can be chosen to be at the right angle to the lateral plane and may be considered as approximately constant at time. Suppose further that an average angle of the in-plane $\vec{k} \to \vec{k}'$ rotation is small ($\vartheta_{\vec{k},\vec{k}'} \ll 1$), then the stochastic scattering process of a wave vector can be modeled by the one-dimensional diffusion in the angular space[5]. These assumptions allow to average Eq.(8) over the stochastic ensemble with the conditional probability density function

$P(\vartheta_{\vec{k},\vec{k}'}, t) = (4\pi D_1 t)^{-1/2} \exp(-\vartheta^2_{\vec{k},\vec{k}'}/4D_1 t)$. Integration over all possible angles $\vartheta_{\vec{k},\vec{k}'}$ gives $<u_{LH,z_L}(t)> = \exp(-D_1 t)$. Consequently, the rate of the geometric spin dephasing in the $LH_1$ band is equal to the rate of the one-dimensional orientational relaxation of the lattice momentum

$$1/T_s = D_1 = 1/\tau_k = <\omega^2>\tau_\omega, \qquad (9)$$

where $<\omega^2>\tau_\omega^2 \ll 1$ and $\tau_\omega$ is the correlation time of $\vec{\omega}$ (see Refs.[10, 30] for details). As long as stochastic rotational perturbation of the Hamiltonian of a system can be considered as adiabatic, this result does not depend on the physical nature of the system. It is of interest to note that Eq.(9) was first derived by Jones and Pines[30], who applied the general method developed in Ref.[29] to describe a geometric dephasing in the context of zero-field NQR experiments of $^{131}$Xe ($I$ = 3/2) *nuclei* induced by thermal collisions with the walls of a toroidal container. This result is only applicable to 2D systems. The 3D case is generally more complex, since $\hat{\vec{n}}$ can change its direction in time, so the elementary rotations in the local and the mesoscopic reference frames may not commute (non-Abelian case), and will be considered elsewhere.

**III. Summary and Discussion.**

It should be clear from the above that the spin dephasing of HH's is expected to be dominated by *nonadiabatic* intersubband $HH_1$-$LH_1$ transitions, which strongly depend on the ratio between $|D_k|$ and $\tau_c$. This process has obviously much smaller rate than the adiabatic (geometric) spin dephasing of LH's, which may qualitatively explain the dramatic increase in the $HH_1$ spin lifetime observed in quantum wells (~1 ns)[32] in comparison to bulk crystals (~ 0.1 ps)[12]. At room temperatures $\tau_c \sim 10^{-13}$ s, the $HH_1$-$LH_1$ splitting in typical 2D nanostructures[33] $\geq$ 50 meV at $\vec{k}=0$, which may not be enough to guarantee the adiabaticity of the carrier's response to the perturbation caused by the hole scattering. However, this is definitely sufficient to significantly suppress the nonadiabatic intersubband transitions and, hence, the spin dephasing in the $HH_1$ band. In this context, it is qualitatively clear why the hole's spin relaxation time was reported to be a rapidly decreasing function of the $HH_1$ population temperature in the GaAs quantum-wells, where merely the $HH_1$ spin quantum beats were observed[34].

Adiabatic suppression of the interband transitions does not mean that $HH_1$ and $LH_1$ doublets become totally decoupled. Similar to fluctuations of a crystal field splitting, the thermal fluctuations of the *magnitude* of an anisotropic part of the instantaneous Hamiltonian will stochastically modulate the gap between the $HH_1$ and the $LH_1$ components of the $\Gamma_8$ quadruplet (parameter $D_k$ is generally time-dependent). This process leads to dephasing of the *dynamic* phase acquired by the components of HH and LH spinors during the collision[35]. To keep the calculations and the results as simple as possible, in this paper we ignore the dynamic dephasing. This approach allows us to focus on the effect of geometric dephasing and pinpoint the particular features of this process. The latter is intrinsically related to the stochastic modulation of the spin-rotation interaction induced by the fluctuations of the Coriolis force that sparkles during elastic collisions (curved trajectories, see also Ref.[24]). Remarkably, within the second order (effective-mass) $\vec{k} \cdot \vec{p}$ perturbation theory the geometric dephasing is singularly ineffective for heavy holes. Since the valence band has *p*-symmetry and does not couple in the first order to nuclear spins, these carriers may be considered as attractive candidates for spintronic devices.